\documentstyle[prb,aps,epsf]{revtex}
\begin{document}

\twocolumn[\hsize\textwidth\columnwidth\hsize\csname
@twocolumnfalse\endcsname

\widetext
\title{Unscreened Coulomb repulsion in the one dimensional 
electron gas} 
\author { G. Fano$^{1}$, F. Ortolani$^{1}$, 
A. Parola$^{2,3}$ and L. Ziosi$^{1}$}
\address{
$^1$ Dipartimento di Fisica, Universit\'a di Bologna, Via Irnerio 46,
40126 Bologna, Italy. \\
$^2$ Istituto Nazionale di Fisica della Materia, Unit\'a di Milano and \\
$^3$Dipartimento di Scienze Fisiche, Universit\'a dell'Insubria, Via
Lucini 3 Como, Italy. }

\maketitle
\begin{abstract}
A tight binding model of electrons interacting via bare Coulomb repulsion
is numerically investigated by use of the Density Matrix Renormalization Group
method which we prove applicable also to very long range potentials. 
From the analysis of the elementary excitations, of the spin and charge
correlation functions and of the momentum distribution, a picture 
consistent with the formation of a one dimensional ``Wigner crystal"
emerges, in quantitative agreement with a previous bosonization study. 
At finite doping, Umklapp scattering is shown to be ineffective in 
the presence of long range forces. 
\end{abstract}
\pacs{ }
]
\narrowtext
One dimensional electron models are often used to interpret the
behavior of strongly anisotropic physical systems in condensed matter, 
like organic conductors \cite{organ}, charge transfer
salts \cite{salt} and certain semiconductor nanostructures \cite{nano}.
These systems can be modeled in terms of a tight binding hamiltonian with 
effective electron repulsion, which can be either short or long
ranged, depending on the extent of interchain screening.
The prototype of the short range models is the 
exactly solvable one dimensional repulsive Hubbard model \cite{liebwu}
which is known to be in the universality class of Luttinger 
liquids \cite{schulzhub},
showing metallic properties and antiferromagnetic 
spin correlations at every finite doping. However, this picture
may dramatically change if screening is not effective in reducing the
range of the bare Coulomb potential. This possibility is well
known in quantum chemistry where the Pariser-Parr-Pople model (PPP),
describing conjugated polyenes \cite{ppp}, has exactly the same structure
of a tight binding model with unscreened Coulomb-like interaction.
This hamiltonian has been studied in small lattices by variational methods, 
like unconstrained Hartree-Fock \cite{hf}, 
suggesting the development of a charge density wave (CDW)
leading to an antiferromagnetic insulating ground state in the
thermodynamic limit. Exact diagonalizations, recently performed 
in systems with few electrons \cite{diago}, instead provide
evidence in favor of a metallic behavior. Unfortunately, 
finite size effects inhibited the study of correlation functions
leaving open the two possibilities of a Luttinger liquid or 
of a CDW metal where the charge carriers can be identified as 
sliding density waves, while electron-like quasiparticles are
absent. The latter scenario was in fact proposed by Schulz \cite{schulz} 
in a seminal bosonization study. Here, the picture of a one dimensional 
Wigner crystal emerged, characterized by extremely long range tails for the
charge correlation functions at the density dependent wavevector $4k_F$, 
together with weaker antiferromagnetic correlations at $2k_F$.

In order to clarify the physics of one dimensional systems
with unscreened Coulomb repulsion, we performed a Density Matrix
Renormalization Group (DMRG) study for the PPP model.
Results for the excitation spectrum in the spin and charge
channels are presented. The density and magnetic structure factors 
have been computed both at half filling and in the doped
system. A physical insight on the nature of charge carriers has
been obtained through the study of the momentum distribution
of the electrons.

DMRG is an extremely accurate numerical method, especially devised for
one dimensional problems, which can easily handle system sizes two
or three times larger than usual diagonalization algorithms, thereby
drastically reducing finite size effects \cite{white}. 
DMRG gives the exact spectrum of the hamiltonian in a reduced
Hilbert space and therefore it
provides a variational bound to the exact ground state 
energy. The method also allows for a self consistent evaluation of
the errors introduced by truncation, by checking the unitarity sum rule
satisfied by the exact density matrix, which reflects the 
completeness of the Hilbert space.

The model we have studied is defined by $N$ electrons on
a $L$ site ring with hamiltonian
\begin{equation}
H=-t\sum_{i,\sigma} \left (c^\dagger_{i,\sigma}c_{i+1,\sigma} +
h.c.\right) + {U\over 2}\sum_{i,j} {(n_i-\bar n) 
(n_j-\bar n)\over 1+\gamma_0\, d_{ij}}
\label{ppp}
\end{equation}
where $c_{i,\sigma}$ is a fermionic annihilation operator, 
$n_i=\sum_{\sigma} c^\dagger_{i,\sigma}c_{i,\sigma}$ is the
density operator at site $i$ with average value
$\bar n=N/L$ and $d_{ij}$ is the chord distance
on the circle:
\begin{equation}
d_{ij}={\left |\sin(i-j){\pi\over L}\right |\over \sin{\pi\over L}}
\label{dij}
\end{equation}
The parameter $\gamma_0$, which controls the strength
of the long range Coulomb repulsion, is
fixed at $\gamma_0=1.053907$, following the Mataga-Nishimoto prescription
\cite{nippo}. Here we present extensive results for a representative value
interaction strength $U/t=13.55$. 

The physics underlying the half filled case (i.e. $\bar n=1$) 
is well established. In the strong
coupling limit (i.e. $U\to \infty$), charges are frozen and the 
low energy configurations just correspond to different spin orientations:
in a bipartite lattice, 
the ground state is always 
a non degenerate spin singlet and
it is expected to show correlations typical of 
one dimensional antiferromagnets: $<{\bf S}_n \cdot {\bf S}_0>\sim
(-1)^n/n$, like in the Heisenberg model. Spin excitations are
gapless with linear dispersion relation.
Analogously to the Hubbard model, for every strength of the
Coulomb repulsion, a charge gap develops, 
the system flows toward strong coupling and its
correlation functions have the same structure as in the $U\to \infty$ limit.
The open questions regard the effects of doping in the presence of
long range repulsive interactions. In the Hubbard model, 
as soon as we insert holes (or extra electrons) into the lattice,
the charge gap closes and both the charge and the spin branch of
the excitation spectrum acquire linear dispersion 
relation, characterized by finite charge and spin velocity.
In this class of spin isotropic Luttinger liquids, the long 
wavelength properties of the system are determined by a single
dimensionless parameter $K_\rho$ governing the power law decay
of all correlation functions.
The antiferromagnetic correlations in the ground state are
characterized by the wavevector $q=2k_F$ and decay as $x^{-(1+K_\rho)}$
while the density correlations are dominated by oscillations at $q=4k_F$
(i.e. wavelength $\lambda=1/\bar n$) and behave as $x^{-4K_\rho}$.
The momentum distribution has no longer a sharp jump at the
electronic Fermi momentum but maintains a power law singularity
of the form $\pm |q-k_F|^\alpha$ with $\alpha=(K_\rho+1/K_\rho-2)/4$.
The analysis based on bosonization techniques \cite{schulz} showed that
the long range nature of Coulomb interaction is expected to change this
picture: while the spin sector is unaffected by interactions, 
the charge spectrum now becomes non analytic: 
$\omega_q\sim q\sqrt{|\ln q |}$.
Formally this corresponds to the $K_\rho\to 0$ limiting case.
In fact, the leading power law behavior of spin correlations is
just $\cos(2k_Fx)/x$ while charge correlations decay more slowly than any
power law and the momentum distribution is continuous with all its
derivatives at the Fermi momentum. Actually, the bosonization
study allowed to determine the precise form of the asymptotic behavior,
which turns out to be:
\begin{eqnarray}
<{\bf S}(x)\cdot {\bf S}(0)> &\sim& \exp\left [-c(\ln x)^{1/2}
\right ]{\cos(2k_Fx)\over x}
\nonumber\\
< n(x) n(0)> &\sim& \exp\left [-4c (\ln x)^{1/2}\right ]\cos(4k_Fx)
\nonumber\\
< c^\dagger_{x,\sigma} c_{0,\sigma}> &\sim& \exp\left [-c^\prime(\ln x)^{3/2}
\right ] \cos(k_Fx)
\label{bose}
\end{eqnarray}
with $c$ and $c^\prime$ non universal positive constants.
However, Umklapp scattering, possibly leading to CDW instabilities,
has been neglected in the bosonization analysis, as remarked in Ref.
\cite{diago}. Other possible scenarios involve charge ordering with 
different periodicity, i.e. the one dimensional analog of a``stripe" phase,
or the formation of a gap in the spin excitation spectrum, as a precursor 
to phase separation, which is inhibited by long range forces.

In order to understand which picture correctly describes the physics
of correlated electron systems with long range interactions,
we numerically studied the PPP model (\ref{ppp}) at the two electron 
densities $\bar n=1/2$ and $\bar n =3/4$ corresponding to $k_F=\pi/4$
and $k_F=3\pi/8$ respectively. These choices of filling factors 
are suggested by the necessity to perform accurate size scaling 
keeping, at the same time, a limited number of degrees of freedom.
Calculations have been performed in lattices up to $L=80$ sites
and total number of electrons up to $N=60$.
The chosen electron densities are represented by small 
fractions, hence we expect that commensurability effects may
be enhanced for these cases: If the model is prone to a CDW instability,
the systems studied in this work should clearly suggest the tendency
toward charge ordering.
Our DMRG code is written in such a way to cope with long range 
potentials. The dimension of the reduced Hilbert space 
is always larger than $10^6$, the truncation error is at
most $10^{-5}$ and the correlation functions are translationally 
invariant up to a relative error of $2\times 10^{-2}$.
Further details on the algorithm can be found in Ref. \cite{ziosi}. 

As a first step, we calculated the charge spectrum, i.e.
the ground state energies $E(N+1)$ and $E(N-1)$ obtained by adding and 
removing one electron to the reference state
with $N=\bar n L$ particles. In order to have a non-degenerate singlet 
ground state, we imposed periodic or anti-periodic boundary conditions
thereby realizing the closed shell condition in the non-interacting 
limit on the reference state.
The finite size gap is defined as the difference between
the upper and lower estimate of the chemical potential:
\begin{equation}
\Delta_\rho=\mu_+-\mu_- = {1\over 2}\left [E(N+1)+E(N-1)-2E(N)\right ]
\label{gap}
\end{equation}
Although $\Delta_\rho$ does not coincide with the true charge gap we can
infer that the charge gap is zero if $\Delta_\rho$ vanishes in the 
thermodynamic limit.
Conversely, the spin gap $\Delta_\sigma$ is just the energy 
difference between the singlet and the triplet spin sectors
at fixed number of particles.
The finite size scaling of our results, shown in Fig. 1,
clearly indicates the gapless nature of the excitation spectrum,
ruling out the possibility of charge ordering in the ground state
and confirming the irrelevance of Coulomb repulsion on the structure of 
spin excitations in agreement with previous diagonalization data
\cite{diago}. The analysis of correlation functions provides
a deeper information on the physical nature of the ground state.
In particular, we studied the charge and spin structure factors
defined by:
\begin{eqnarray}
R(q)&=&={1\over L}\sum_{l,m}e^{iq(l-m)} <(n_l-\bar n)\,(n_m-\bar n)>
\nonumber\\
S(q)&=&={1\over L}\sum_{l,m}e^{iq(l-m)} <S^z_l\,S^z_m>
\label{defcor}
\end{eqnarray}
The Fourier transform $R(q)$ of the equal time
density-density correlation function shown in Fig. 2
displays a remarkable collapse
of data relative to different sizes, with the single exception of
the peak region at $q=4k_F$. The peak value, plotted as a function
of $L$ in Fig. 3, is well fitted by the expected form $R(4k_f)\sim
L\exp(-4c\sqrt{\ln L})$ deduced from the bosonization analysis (\ref{bose})
suggesting that quasi long range order develops in the system.
Notice the appreciable curvature of the data which reflects 
the presence of the exponential term in the fitting formula and,
in turn, indicates a very slow decay of density 
correlations
in real space. The holes effectively repel each other and stay almost rigidly 
at the maximum attainable average distance $\lambda=1/\bar n$
but, as expected, hole correlations vanish at large
distance, in agreement with Eq. (\ref{bose}).
A much weaker singularity seems to be present at wavevector $2k_F$,
but the data do not allow for a systematic analysis of this further feature,
also predicted by the bosonization analysis of Ref. \cite{schulz}.
Other interesting information can be extracted from the small $q$
behavior of $R(q)$ which is related to the low energy
excitation spectrum. According to the bosonization formulas, the 
structure factor should behave as
$R(q)\sim q | \ln q |^{-1/2}$. This form well represents the
DMRG data, as shown in Fig. 4. Note that such an expression is quite specific
to the bare Coulomb interaction and differs from the 
usual Luttinger liquid results, where $R(q)\sim K_\rho (q/\pi) $.
A similar analysis can be carried out for the magnetic properties of the
model. In Fig. 5 the magnetic structure 
factors $S(q)$ corresponding to
the previously discussed choices of parameters are shown. The collapse
of points on the same smooth curve should be again appreciated. The dominant
peak occurs at $q=2k_F$, as expected, and its height, reported in Fig. 6, 
scales as $a_0-a_1(\sqrt{\ln L}+1/c)\exp(-c\sqrt{\ln L})$
in agreement with the bosonization prediction (\ref{bose}). 
Here $a_0$ and $a_1$ are fitting parameters, while the constant $c$  
is fixed to the same value obtained from the fit of the density
structure factor shown in Fig. 3. Contrary to the results obtained for
$R(q)$, the small $q$ behavior of $S(q)$
is accurately given by the linear relation $S(q)\sim (q/4\pi)$ 
corresponding to $K_\sigma=1$, as expected for all the gapless,
$SU(2)$ invariant, one dimensional electron systems.

The previously discussed properties of the PPP model, emerging
from the DMRG analysis, are consistent with the Wigner crystal model
suggested by the bosonization study: 
away from half filling, the system is gapless
both in the charge and spin sectors but, at the same time, it supports 
extremely long ranged density correlations. Following Ref. \cite{diago},
we may consider two different pictures, consistent with the observed metallic
properties of this model. The first one corresponds to a Luttinger liquid 
scenario, where the electrons have a sort of ``Fermi surface" 
satisfying the Luttinger theorem. In this case, although the
low energy excitations are always collective in one dimension, 
the charge carriers in the system may be still identified with dressed 
electrons. Instead, according to the second picture, 
the absence of a gap in the charge spectrum 
is due to the Wigner crystal nature of the ground state and to the 
long range decay of density correlations. The quasiparticles cannot 
be identified with electrons any more and consequently the electron momentum
distribution is smooth at the Fermi momentum, with an extremely weak
essential singularity, as reported in Eq. (\ref{bose}).
DMRG data for the momentum distribution for given spin $n(q)$ 
are shown in Fig. 7, while a
size scaling of the finite size jump at $q=k_F$ is plotted in Fig. 8. 
Again, the asymptotic scaling 
\begin{equation}
\Delta n(k_F)\sim Le^{-c^\prime (\ln L)^{3/2}}
\label{jump}
\end{equation}
predicted on the basis of the bosonization result (\ref{bose}) gives
a quite good representation of numerical data.

In conclusion, we have tried to clarify  the physics 
underlying the presence of unscreened Coulomb interaction in one
dimension. Accurate DMRG data are fully consistent with the
picture of a Wigner crystal showing metallic properties, with 
slowly decaying density correlation and gapless charge and spin
excitation spectra. At least in one dimension, this numerical
investigation rules out other scenarios, like pinned CDW or
even some precursor of off-diagonal long range order.

\clearpage
\begin{figure}
\centerline{
\epsfxsize = 8.6truecm\epsfbox{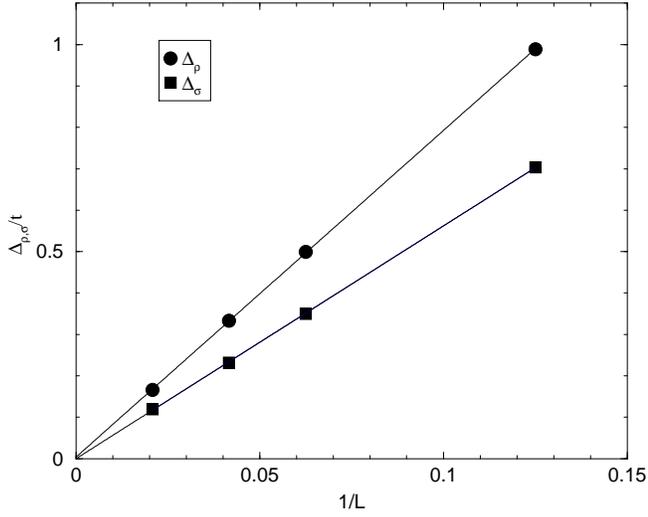}}
\caption{
Charge gap $\Delta_\rho$ and spin gap $\Delta_\sigma$, 
in units of $t$, versus the 
inverse chain length $L$. The density is $\bar n = 3/4$.
\label{fig1}}
\end{figure}

\begin{figure}
\centerline{
\epsfxsize = 8.6truecm\epsfbox{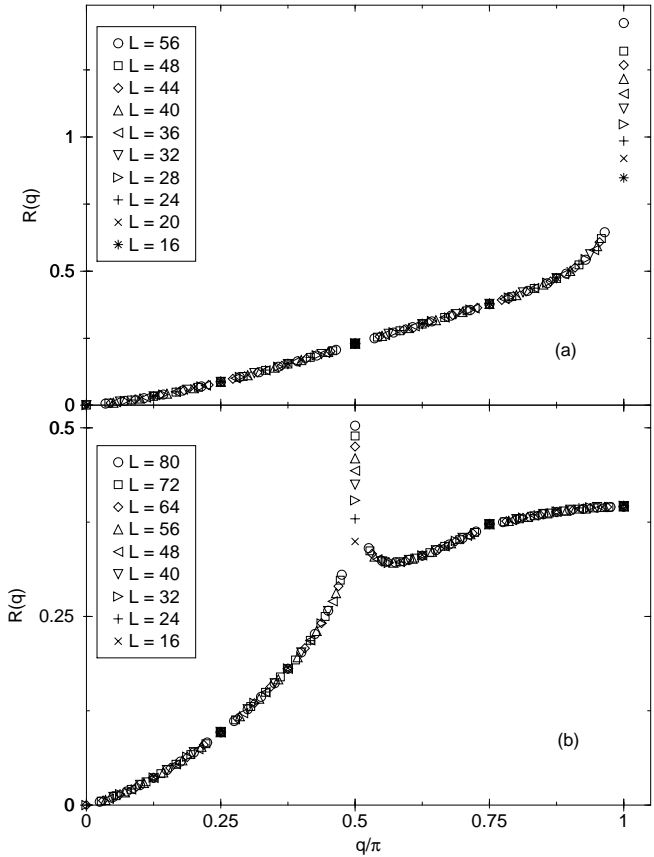}}
\caption{
FIG. 2. Fourier transform $R(q)$ of the equal time density-density 
correlation function. The density is $\bar n = 1/2$ in (a),
$\bar n = 3/4$ in (b).
\label{fig2}}
\end{figure}

\begin{figure}
\centerline{
\epsfxsize = 8.6truecm\epsfbox{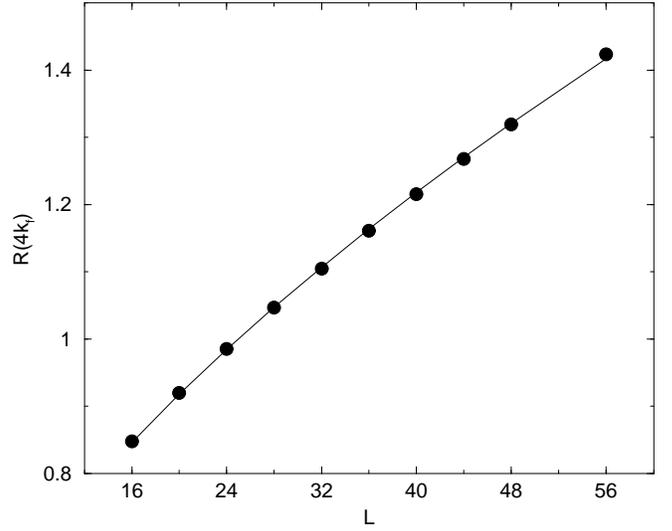}}
\caption{
Fit of $R(4k_f)$ with respect to the bosonization result.
The circles are DMRG results for density $\bar n=1/2$ 
and the solid line is
a function of the form 
$a L \exp(-4c\sqrt{\ln{L}})$ 
with $a=1.93$ and
$c=0.54$.
\label{fig3}}
\end{figure}

\begin{figure}
\centerline{
\epsfxsize = 8.6truecm\epsfbox{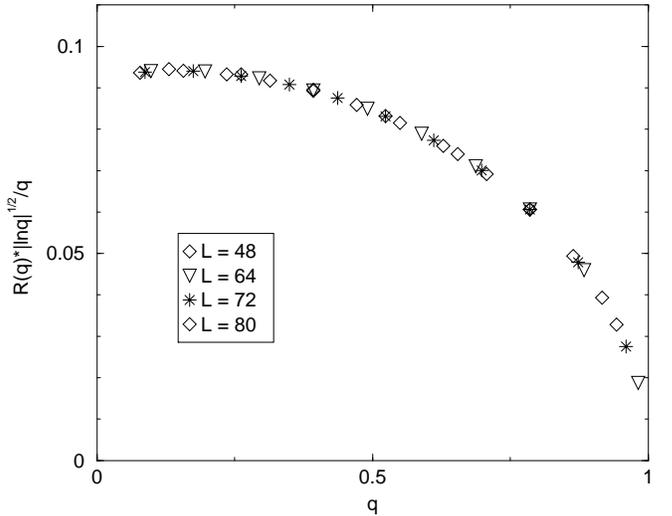}}
\caption{
Small momentum behavior of the charge structure factor. We plot R(q)  
divided by the expected limiting form $q|\ln(q)|^{-1/2}$.
The density is $\bar n = 3/4$.
\label{fig4}}
\end{figure}

\begin{figure}
\centerline{
\epsfxsize = 8.6truecm\epsfbox{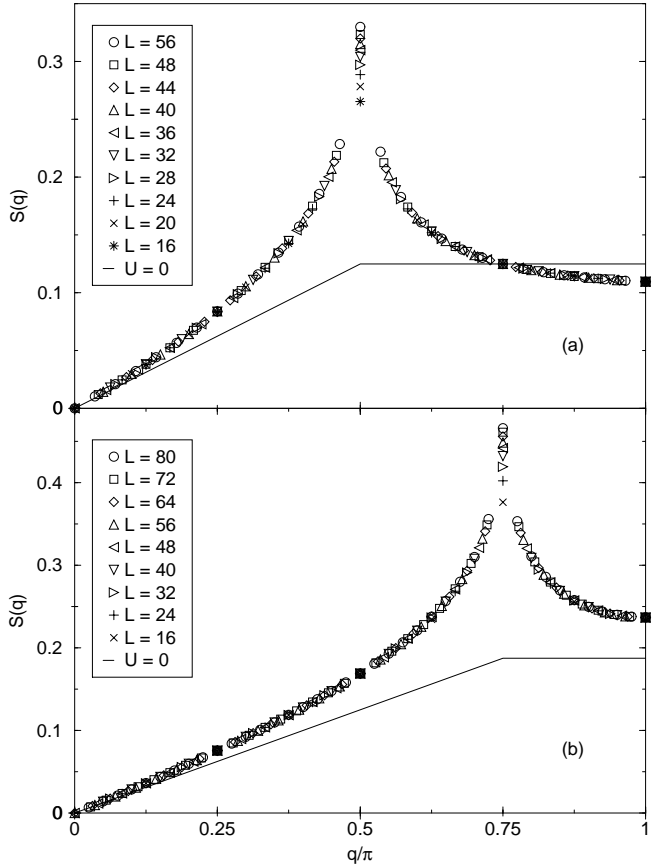}}
\caption{
\
Fourier transform S(q) of the equal time spin-spin 
correlation function. The density is $\bar n = 1/2$ in (a) and 
$\bar n = 3/4$ in (b).
\label{fig5}}
\end{figure}

\begin{figure}
\centerline{
\epsfxsize = 8.6truecm\epsfbox{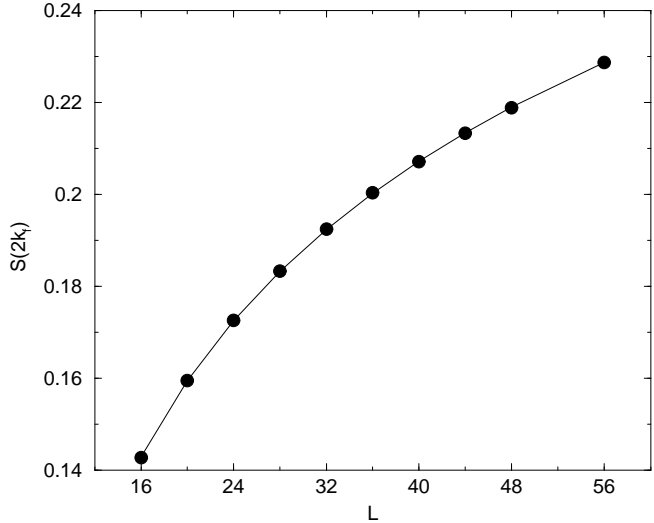}}
\caption{
Fit of $S(2k_f)$ with respect to the bosonization result.
The circles are DMRG results for $\bar n=1/2$ and the solid line is 
a function of the form $a_0-a_1(\sqrt{\ln{L}}+1/c)\exp(-c\sqrt{\ln{L}})$ 
with $a_0=1.12$, $a_1=0.68$ and $c=0.54$ (obtained from Fig. 3).
\label{fig6}}
\end{figure}

\begin{figure}
\centerline{
\epsfxsize = 8.6truecm\epsfbox{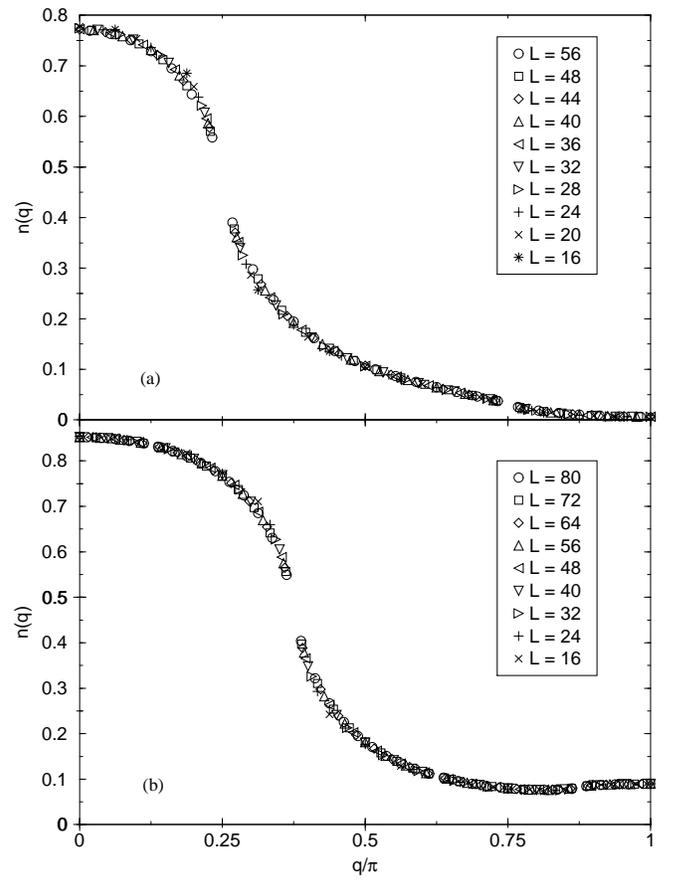}}
\caption{
Momentum distribution function for given spin $n(q)$.
The density is $\bar n =1/2$ in (a) and $\bar n=3/4$ in (b). 
\label{fig7}}
\end{figure}

\begin{figure}
\centerline{
\epsfxsize = 8.6truecm\epsfbox{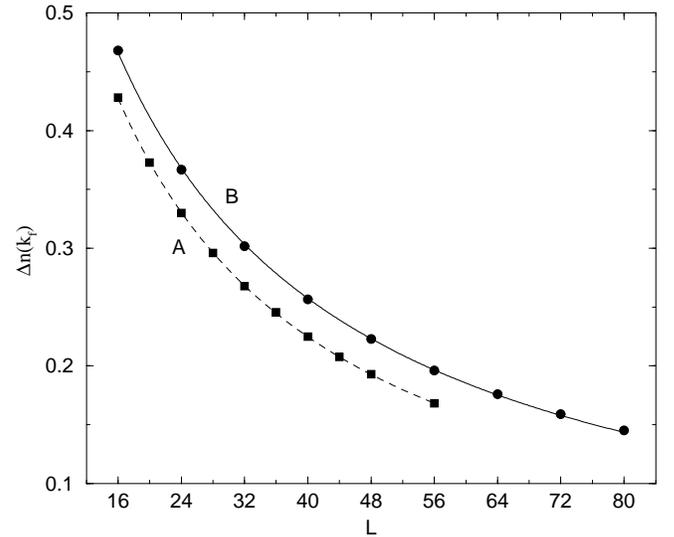}}
\caption{
Fit of the finite size jump at $k_f$ of the momentum distribution
with the expression: $\Delta n(k_f) = a L e^{-c'\ln(L)^{3/2}}$.
The parameters are: $a = 0.493781$, $c' = 0.631772$ for $\bar n = 1/2$
(curve $A$) and $a=0.491581$, $c' = 0.611964$ for $\bar n = 3/4$ (curve $B$).
\label{fig8}}
\end{figure}

\end{document}